%% file: content/Main.tex
\documentclass[sigconf]{acmart}
\usepackage{listings}
\usepackage{xcolor}
\usepackage{caption}
\usepackage{tabularx}
\usepackage{booktabs}
\usepackage{makecell}

\lstdefinestyle{plaintext}{
  basicstyle=\ttfamily\small,
  breaklines=true,
  frame=single,
  backgroundcolor=\color{gray!5},
  columns=fullflexible
}

\AtBeginDocument{%
  }

\setcopyright{acmlicensed}
\copyrightyear{2018}
\acmYear{2018}
\acmDOI{XXXXXXX.XXXXXXX}
\acmConference[Conference acronym 'XX]{Make sure to enter the correct
  conference title from your rights confirmation email}{June 03--05,
  2018}{Woodstock, NY}
\acmISBN{978-1-4503-XXXX-X/2018/06}

\acmVolume{37}
\acmNumber{4}
\acmArticle{111}
\acmMonth{8}

\renewcommand\footnotetextcopyrightpermission[1]{} % Removes the copyright footer
\makeatletter
\renewcommand\@formatdoi[1]{} % Removes the DOI
\makeatother

\begin{document}

%%
%% The "title" command has an optional parameter,
%% allowing the author to define a "short title" to be used in page headers.
% \title{Domain-Driven Adaptable AI Pipelines}
% \title{AI to Build AI: Leveraging Models to Guide Researchers in Problem Solving}
% \title{Lowering the Entry Barrier: Leveraging AI Models to Guide Researchers in Problem Solving}
% \title{From Intent to Implementation: A Controlled Agentic-AI Framework for Non-Experts}
% \title{AI Pipeline Generation for Scientists Without AI Expertise Using a Controlled Agentic Framework}
\title{From Intent to AI Pipelines: A Controlled Agentic Framework for Non-AI Expert Scientists}

% \title{Lowering the Entry Barrier: Domain-Driven AI Solutions}
% \title{Lowering the Entry Barrier: Leveraging Models to Guide Researchers in Problem Solving}

%%
%% The "author" command and its associated commands are used to define
%% the authors and their affiliations.
%% Of note is the shared affiliation of the first two authors, and the
%% "authornote" and "authornotemark" commands
%% used to denote shared contribution to the research.

\author{Hyacinth Ali}
\affiliation{%
  \institution{University of Montreal}
  \city{Montreal}
  \country{Canada}}
\email{hyacinth.chijioke.ali@umontreal.ca}

\author{Jessie Galasso-Carbonnel}
\affiliation{%
 \institution{McGill University}
 \city{Montreal}
 \country{Canada}}
 \email{jessie.galasso-carbonnel@mcgill.ca}

\author{Houari Sahraoui}
\affiliation{%
  \institution{University of Montreal}
  \city{Montreal}
  \country{Canada}}
  \email{sahraouh@iro.umontreal.ca}

\pagestyle{plain} % This replaces the conference headers with simple page numbers

%%
%% By default, the full list of authors will be used in the page
%% headers. Often, this list is too long, and will overlap
%% other information printed in the page headers. This command allows
%% the author to define a more concise list
%% of authors' names for this purpose.
\renewcommand{\shortauthors}{Ali et al.}

%%
%% The abstract is a short summary of the work to be presented in the
%% article.
\begin{abstract}
Artificial Intelligence (AI) pipelines have become integral to modern research, supporting fields such as Medical Sciences, Agriculture, and Social Sciences, and enabling large-scale data analysis, predictive modeling, and the automation of complex tasks. However, designing and implementing AI solutions remains challenging for many researchers due to the expertise required in the design and development of end-to-end AI systems. To address this gap, we present Domain-Driven Adaptable AI Pipelines (DDAP), a controlled, human-in-the-loop, agentic framework that leverages large language models to guide users in a systematic construction of AI pipelines and their corresponding implementation code. \emph{DDAP} structures the development process into four stages: problem definition, compute environment specification, pipeline generation, and code generation. Through this staged interaction, the framework adapts to domain context, user expertise, and resource constraints, while maintaining user control over key decisions. We evaluate \emph{DDAP} across multiple datasets spanning business, biology, and health science domains by comparing its AI models against expert-developed models. The experimental results show that DDAP achieves competitive results in several tasks compared to expert baselines, although performance varies across problem types, particularly for text-based clustering tasks. By combining guided interaction, adaptability, and reproducibility, \emph{DDAP} demonstrates that a controlled agentic framework can generate competitive AI pipelines for non-expert users.

\end{abstract}

%%
%% The code below is generated by the tool at http://dl.acm.org/ccs.cfm.
%% Please copy and paste the code instead of the example below.
%%
% \begin{CCSXML}
% <ccs2012>
%  <concept>
%   <concept_id>00000000.0000000.0000000</concept_id>
%   <concept_desc>Do Not Use This Code, Generate the Correct Terms for Your Paper</concept_desc>
%   <concept_significance>500</concept_significance>
%  </concept>
%  <concept>
%   <concept_id>00000000.00000000.00000000</concept_id>
%   <concept_desc>Do Not Use This Code, Generate the Correct Terms for Your Paper</concept_desc>
%   <concept_significance>300</concept_significance>
%  </concept>
%  <concept>
%   <concept_id>00000000.00000000.00000000</concept_id>
%   <concept_desc>Do Not Use This Code, Generate the Correct Terms for Your Paper</concept_desc>
%   <concept_significance>100</concept_significance>
%  </concept>
%  <concept>
%   <concept_id>00000000.00000000.00000000</concept_id>
%   <concept_desc>Do Not Use This Code, Generate the Correct Terms for Your Paper</concept_desc>
%   <concept_significance>100</concept_significance>
%  </concept>
% </ccs2012>
% \end{CCSXML}



% \ccsdesc[500]{Do Not Use This Code~Generate the Correct Terms for Your Paper}
% \ccsdesc[300]{Do Not Use This Code~Generate the Correct Terms for Your Paper}
% \ccsdesc{Do Not Use This Code~Generate the Correct Terms for Your Paper}
% \ccsdesc[100]{Do Not Use This Code~Generate the Correct Terms for Your Paper}

%%
%% Keywords. The author(s) should pick words that accurately describe
%% the work being presented. Separate the keywords with commas.

% \received{20 February 2007}
% \received[revised]{12 March 2009}
% \received[accepted]{5 June 2009}

\maketitle

\input{content/1-Introduction}

\input{content/2-Background}

\input{content/3-DDAP_Framework}

\input{content/4-DDAP_Experiments}

\input{content/5-Related_Work}

\input{content/6-Conclusion}

\newpage
\bibliographystyle{ACM-Reference-Format}
\bibliography{content/low-code}

\end{document}

%% file: content/1-Introduction.tex
\section{Introduction}
Artificial Intelligence (AI) has become a cornerstone of modern scientific research, with applications spanning healthcare \cite{lee2021application}, natural sciences \cite{khan2023future}, and social sciences \cite{goyanes2025use}. Increasingly, scientists leverage AI pipelines to analyze complex datasets, automate experimentation, and accelerate discovery. This widespread adoption signals a transformative shift in how research is conducted across disciplines. 
Despite these advances, scientists face significant challenges when attempting to design and implement AI pipelines independently. Developing an effective AI solution requires expertise at multiple levels: formulating a research question as an AI task, preparing data, selecting and training/aligning models, and finally deploying solutions in real-world environments. Each of these stages presents non-trivial barriers, particularly for researchers with limited backgrounds in machine learning or software engineering \cite{nair2024barriers,reddy2025generative}. Consequently, many scientists remain dependent on collaborations with AI specialists, a dependency that can slow down research progress and limit accessibility \cite{goyanes2025use}.

A variety of low-code and no-code frameworks have been proposed to address these barriers \cite{shyam2025bridging,maddireddy2025locoml}. While such frameworks lower the technical entry point, they often lack sufficient flexibility, adaptability to domain-specific constraints, or support for the entire pipeline lifecycle. More recently, large language models (LLMs) have been used directly by scientists to generate pipeline code from natural language descriptions. Although promising, this approach faces important limitations. First, LLM outputs can be unreliable, producing incomplete, non-reproducible, or suboptimal pipelines. Second, LLMs frequently overlook domain-specific constraints, such as regulatory requirements in healthcare or experimental reproducibility in the natural sciences. Third, effective use of LLMs still demands significant technical skills: crafting appropriate prompts, validating generated code, and debugging errors all place a high cognitive load on non-experts. Finally, the opaque decision-making of LLMs introduces challenges for transparency and reproducibility \cite{arslan2024survey,ayyamperumal2024current}.

These issues point to a broader methodological gap: while frameworks exist for conceptualizing the transition from problem descriptions to machine learning workflows \cite{oakes2024building}, existing tools do not sufficiently operationalize this process in a way that is accessible to non-experts. This gap motivates the need for a more guided and domain-sensitive approach to AI pipeline design, one that integrates domain expertise directly into pipeline construction while lowering technical barriers.

To address these challenges, we frame AI pipeline construction as a software engineering problem of synthesizing executable workflows from high-level domain intent under uncertainty. The key difficulty lies not only in model selection, but in systematically transforming incomplete, domain-specific problem descriptions into consistent, reproducible, and executable software artifacts.

We propose Domain-Driven Adaptable AI Pipelines (DDAP), a controlled, human-in-the-loop, agentic framework that operationalizes this process as an artifact-driven workflow. Rather than treating large language models as fully autonomous generators, DDAP organizes them as coordinated agents within a staged orchestration process that incrementally refines problem specifications into structured pipeline designs and implementation code. This design explicitly emphasizes artifact generation, traceability, and reuse, aligning AI pipeline construction with established software engineering principles such as modularity, separation of concerns, and reproducibility.

% To address these challenges, we propose \emph{Domain-Driven Adaptable AI Pipelines} (DDAP), a controlled agentic framework that leverages large language models as guided assistants within a structured workflow. Rather than treating LLMs as fully autonomous generators, \emph{DDAP} orchestrates their use through a four-stage process: \emph{problem definition}, \emph{compute environment specification}, \emph{pipeline generation}, and \emph{code generation}. This staged interaction is positioned to allow users to remain in control of key decisions while benefiting from automated guidance. The framework adapts to domain context, user expertise, and available computational resources, producing structured artifacts and executable implementations that can be reused across projects.

\paragraph{Contributions.}
This paper makes the following contributions:

\begin{itemize}
    \item \textbf{An artifact-driven formulation of AI pipeline generation as a software engineering problem.}
    We model pipeline construction as a staged synthesis process that transforms high-level domain intent into executable artifacts, moving beyond AutoML approaches that rely on search over predefined components.

    \item \textbf{A controlled agentic framework for end-to-end pipeline synthesis.}
    Unlike low-code/no-code platforms based on static templates, DDAP dynamically adapts to domain context, user expertise, and resource constraints through guided interaction and staged refinement.

    \item \textbf{An orchestrated multi-stage workflow for reliable LLM-based generation.}
    In contrast to single-step LLM prompting, \emph{DDAP} decomposes pipeline generation into coordinated stages with explicit artifact reuse and control logic, improving reproducibility and user control.

    \item \textbf{An empirical evaluation demonstrating competitive performance without iterative optimization.}
    Across multiple domains, DDAP generates pipelines in a single pass that achieve performance comparable to expert-designed models, while exposing limitations in complex tasks such as textual clustering.
\end{itemize}

% This paper makes the following contributions:
% \begin{itemize}
%     \item We introduce the \emph{DDAP} framework, a dynamic human-in-the-loop framework that can guide researchers through the full lifecycle of AI pipeline development using large language models.
    
%     \item We design a domain-driven workflow covering \emph{problem definition}, \emph{compute environment specification}, \emph{pipeline generation}, and \emph{code generation}, which aims to support reproducible and transferable AI development artifacts.
    
%     \item We leverage large language models to generate structured pipeline artifacts and implementation code while adapting to domain context, user expertise, and resource constraints.
    
%     \item We evaluate \emph{DDAP} across multiple datasets and domains, showing that pipelines generated in a single pass, i.e., without iterative hyperparameter tuning, can achieve competitive performance relative to expert-developed models, with variation depending on the task.
% \end{itemize}

The remainder of this paper is structured as follows. Section~\ref{sec:background} presents the background of the study. Section~\ref{framework-architecture} introduces the DDAP framework and its components. Section~\ref{sec:experiments} presents several experiments, illustrating the applicability of the \emph{DDAP} framework across different domains. Section~\ref{sec:related} discusses related work, and Section~\ref{sec:conclusion} concludes the paper.

%% file: content/2-Background.tex
\section{Background}\label{sec:background}
The section presents the background necessary to support the proposed controlled agentic framework. We first review Large Language Models (LLMs) and discuss techniques for improving their reliability, controllability, and task performance. These techniques are central to the orchestration mechanisms used in the \emph{DDAP} framework. We then introduce the evaluation metrics employed in this work to assess the effectiveness of the generated AI pipelines across classification, regression, and clustering tasks.
    
    \subsection{Large Language Model}
    Large Language Models (LLMs) are foundation models~\cite{bommasani2021opportunities} based on deep neural networks and Transformer architecture. LLMs have demonstrated outstanding capabilities across a wide range of Natural Language Processing (NLP) tasks~\cite{minaee2024large}, as well as code generation~\cite{joel2024survey}. These capabilities position LLMs as foundational components for building intelligent systems that can assist users in problem formulation, reasoning, and solution generation~\cite{gu2025large}.

    Despite these strengths, LLMs possess several limitations, particularly, training and deployment of LLMs require substantial computational resources, which limits accessibility for many users and organizations. Moreover, LLMs are prone to generating plausible yet incorrect outputs, which raises concerns about reliability in high-stakes and safety-critical systems. These challenges highlight the need for structured frameworks that can harness the capabilities of LLMs while mitigating their limitations, an objective that motivates their guided integration within the proposed \emph{DDAP} framework.

    \subsection{LLM Adaptation and Prompt Techniques}
    To improve the performance of LLMs, AI practitioners usually have two broad options: (1) Fine-Tuning, and (2) Prompt Engineering. Fine-tuning a Large Language Model (LLM) involves adjusting a subset of its parameters to align its performance with a specific task and dataset~\cite{min2023recent}. Although the fine-tuning approach can improve the model’s ability in a particular domain or task, it presents two major limitations: (1) it often requires substantial computational resources; and (2) many tasks lack sufficient labelled data to support effective fine-tuning. In this regard, LLM outputs need to be guided towards desired outcomes, i.e., prompting. The prompt-based technique relies on carefully designed input message~\cite{liu2023pre}. There are many types of prompt techniques, and we briefly present a few of them in the sub-sections below.

        \subsection{System Message}
        System message prompting refers to the use of structured, persistent instructions that define the behaviour, objectives, and constraints of a Large Language Model (LLM) throughout an interaction. Unlike user prompts, which are typically task-specific and transient, system messages establish a global context that guides the model’s behaviour across multiple steps \cite{liu2023pre,shanahan2023role}. 
    
        \subsubsection{Role Based}
        Role-based prompting involves assigning a specific role or persona to an LLM, such as a  \emph{professor} or \emph{ML Engineer}, to guide its reasoning and response style~\cite{shanahan2023role}. By conditioning the model to operate within a defined context, this approach helps align outputs with domain-specific expectations and reduces ambiguity.

        % In the \emph{DDAP} framework, role-based prompting is used to dynamically adapt the framework’s behaviour across different stages of the pipeline, e.g., as a problem-definition assistant, pipeline designer, or code generator, aiming that the generated outputs remain consistent with the intended task and user expertise level.
        
        \subsubsection{Guardrails}
        Guardrails refer to mechanisms that constrain LLM outputs to ensure adherence to predefined requirements~\cite{ayyamperumal2024current, rebedea2023nemo}. These constraints can take the form of structured instructions, validation checks, or controlled interaction flows that prevent the model from producing incomplete or irrelevant responses. Guardrails are often used to enforce stepwise progression within a given system message, such that important information needed for that interaction is well captured chronologically.
        
        \subsubsection{Chain-of-Thought}
        Chain-of-Thought (CoT) prompting encourages LLMs to explicitly articulate intermediate reasoning steps when solving complex problems~\cite{wei2022chain}. By decomposing tasks into smaller, sequential steps, \emph{CoT} improves both the accuracy and transparency of model outputs, which is particularly beneficial in complex tasks such as generating AI pipelines and implementation codes. 

    \subsection{Retrieval-Augmented Generation (RAG)} 
     Retrieval-Augmented Generation is a mechanism that enhances the capabilities of Large LLMs by integrating external knowledge sources into the generation process~\cite{arslan2024survey, lewis2020retrieval}. This approach combines the strengths of retrieval-based and generative models, empowering chatbots to have a more nuanced understanding of user intent and the potential to deliver tailored, well-informed outputs~\cite{d2024novel}. 
     
     % The RAG method aims to improve factual grounding, reduce hallucinations, and enable models to operate effectively in domain-specific or knowledge-intensive tasks.

    % Given a user query, a RAG-based system searches for relevant information from external sources, e.g., a dataset, configuration, or specification. To this effect, the system augments users' messages with the retrieved information, which is fed into an LLM. The model subsequently synthesizes a response by leveraging the users' message and the retrieved contextual information, improving factual accuracy and contextual alignment. RAG is particularly effective in conversational systems, as it enables the handling of complex, knowledge-intensive queries, supports access to up-to-date information, and facilitates reasoning over external data sources, ultimately enhancing the precision and relevance of generated responses.

    \subsection{AI Metric Evaluation}
    Evaluating the performance of machine learning models is a critical step in the development of AI systems, as it provides quantitative measures of how well a model achieves its intended objectives. The choice of evaluation metrics depends on the nature of the task, such as classification, regression, or clustering, and directly influences model selection and comparison. 

    \begin{table}[h]
    \centering
    \caption{AI model predictions.}
    \label{tab:classification_outcomes}
    \renewcommand{\arraystretch}{1.2}
    \begin{tabular}{|l|c|c|}
    \hline
    & \textbf{Predicted Positive} & \textbf{Predicted Negative} \\
    \hline
    \textbf{Actual Positive} & True Positive (TP) & False Negative (FN) \\
    \textbf{Actual Negative} & False Positive (FP) & True Negative (TN) \\
    \hline
    \end{tabular}
    \end{table}
    
        \subsubsection{Accuracy}
        Accuracy measures the proportion of correctly predicted instances, whether positive or negative, of the total number of predictions. However, accuracy can be misleading in imbalanced datasets, where high performance on the majority class may obscure poor performance on minority classes~\cite{ghanemlimitations}. As such, it is often complemented by additional metrics such as precision and recall. Referring to the AI model predictions summarized in Table~\ref{tab:classification_outcomes}, accuracy is defined as: 

        \begin{equation}
            \text{Accuracy} = \frac{TP + TN}{TP + TN + FP + FN}
            \label{eq:accuracy}
        \end{equation}
            
        \subsubsection{Precision}
        Precision measures the proportion of true positive predictions over all predicted positive outcomes. It reflects the ability of an AI model to avoid false positives, which is particularly important in applications where incorrect positive predictions carry a high cost, such as medical diagnosis or fraud detection. The precision of an AI model is expressed as:

        \begin{equation}
            \text{Precision} = \frac{TP}{TP + FP} 
            \label{eq:precision}
        \end{equation}
        \subsubsection{Recall}
        Recall measures the proportion of true positive predictions over the total actual positive instances. It evaluates the ability of the model to capture all relevant instances, which is critical in scenarios where missing positive cases are undesirable, such as disease detection. Recall is expressed as:
        \begin{equation}
            \text{Recall} = \frac{TP}{TP + FN}
            \label{eq:recall}
        \end{equation}

        \subsubsection{F1 Score}
        The F1 score is the harmonic mean of precision and recall, providing a balanced measure that accounts for both false positives and false negatives. It is especially useful when dealing with imbalanced datasets, as it offers a more comprehensive view of model performance, and it is expressed as:

        \begin{equation}
            \text{F1 Score} = \frac{2 \times \text{Precision} \times \text{Recall}}{\text{Precision} + \text{Recall}}
            \label{eq:f1-score}
        \end{equation}
        \subsubsection{Mean Absolute Error}
        Mean Absolute Error (MAE) is widely used in regression tasks to measure the average magnitude of prediction errors by averaging the absolute differences between predicted and true values. A perfect model will have an \emph{MAE} of zero (0). It provides an intuitive interpretation of how far predictions deviate from actual values, and the lower the \emph{MAE} value, the better the model performance. The \emph{MAE} is expressed as:

        \begin{equation}
            \text{MAE} = \frac{1}{n} \sum_{i=1}^{n} \left| y_i - \hat{y}_i \right|
            \label{eq:mae}
        \end{equation}
        Where: (i) $n$ is the total number of samples, (i)$y_i$ is the true value of the $i$-th sample, and (iii) $\hat{y}_i$ is the predicted value of the $i$-th sample.

        \subsubsection{Silhouette Score}
        The silhouette score is a clustering evaluation metric that assesses how well data points are grouped within clusters. It measures the similarity of an instance to its own cluster compared to other clusters, with higher values indicating better-defined and more distinct clusters. It is expressed as:

        \begin{equation}
            s(i) = \frac{b(i) - a(i)}{\max\{a(i),\, b(i)\}}
            \label{eq:silhouette}
        \end{equation}
        Where: (i) $s(i)$ is the silhouette score of sample $i$, (i) $a(i)$ is the average distance between sample $i$ and all other points in the same cluster, and (iii) $b(i)$ is the minimum average distance between sample $i$ and all points in any other cluster.

% In summary, while Large Language Models offer strong capabilities for reasoning, code generation, and workflow automation, their effective application requires structured mechanisms to ensure reliability, controllability, and improved performance. Techniques such as system message prompting, guardrails, chain-of-thought reasoning, and retrieval-augmented generation provide complementary approaches for guiding LLM behavior in complex systems. In addition, well-defined evaluation metrics are essential for rigorously assessing the performance of AI models across classification, regression, and clustering settings. The next section presents the \emph{DDAP} framework, which operationalizes these techniques within a controlled, human-in-the-loop, agentic workflow. 

%% file: content/3-DDAP_Framework.tex
\section{DDAP Framework}\label{framework-architecture}
This section introduces the \textbf{D}omain-\textbf{D}riven \textbf{A}daptable \textbf{A}I \textbf{P}ipelines (DDAP) framework. \emph{DDAP} is a controlled, human-in-the-loop, agentic AI framework that orchestrates the use of LLMs within a structured workflow, rather than relying on fully autonomous generation. This section provides an overview of the proposed approach, its formalization, followed by a detailed description of the framework’s architecture and workflow.

\begin{figure}
    \centering
    \fbox{\includegraphics[width=0.95\linewidth]{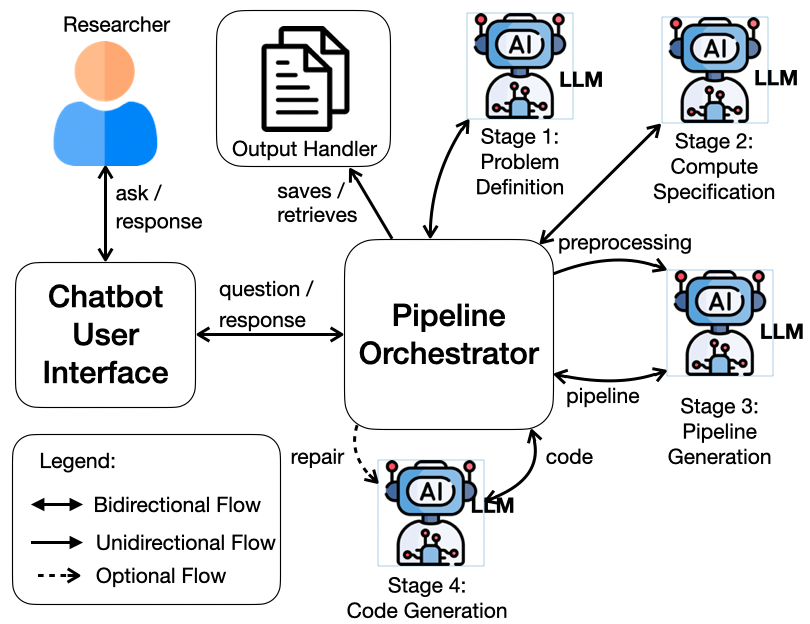}}
    \caption{Domain-Driven AI Pipelines Architecture}
    \label{fig:ddap-architecture}
\end{figure}

\subsection{Overview}

The \emph{DDAP} framework operationalizes a controlled, human-in-the-loop, agentic workflow. 
%As formalized in Section~\ref{sec:Formalization}
This workflow is structured as a sequence of artifact transformations orchestrated across multiple stages. At a high level, DDAP follows an intent-to-implementation paradigm, where users express their problem in domain-specific language, and the system incrementally refines this input into structured artifacts and executable AI pipelines.

Rather than treating Large Language Models (LLMs) as fully autonomous generators, DDAP organizes them as specialized agents coordinated by a central pipeline orchestrator. This design promotes controlled progression across stages while maintaining alignment with user intent, domain constraints, and available computational resources. By decomposing pipeline construction into multiple stages, DDAP reduces the complexity of end-to-end generation and supports more reliable and interpretable outputs.

The architecture of the \emph{DDAP} framework, shown in Figure~\ref{fig:ddap-architecture}, comprises three key components: (1) a chat-based user interface that captures domain intent through natural language interaction, (2) a pipeline orchestrator that manages workflow progression and maintains control logic, and (3) a set of role-based LLM agents responsible for stage-specific tasks such as problem definition, pipeline design, and code generation. Unlike monolithic LLM-based systems~\cite{d2024novel}, this modular design fosters explicit control over each stage of the pipeline development process.

To guide users in developing AI solutions, \emph{DDAP} structures interaction into four stages: \emph{problem definition}, \emph{compute environment specification}, \emph{pipeline generation}, and \emph{code generation}. Each stage produces a structured artifact that is persisted using the \emph{Output Handler} and reused in subsequent stages. Formally, a set of artifacts produced during a pipeline construction is expressed as:
\begin{equation}
    \mathcal{A} = \{A_1, A_2, A_3, A_4\}
    \label{eq:artifcat_definition}
\end{equation}
Where: (i) $A_1$ is the problem definition, (ii) $A_2$ is the compute environment specification, (iii) $A_3$ is the pipeline specification, and (iv) $A_4$ is the executable code. Each artifact captures a progressively refined representation of the initial user intent. For example, the problem definition artifact captures task objectives, data characteristics, and constraints, which subsequently inform both the compute environment specification and pipeline generation stages. This artifact-driven design promotes reproducibility, traceability, and cross-project reuse.

Each agent is implemented using a large language model (LLM) configured with role-based instructions, guardrails, and contextual prompts. Formally, a set of agents is expressed as:
\begin{equation}
    \mathcal{G} = \{\alpha_1, \alpha_2, \alpha_3, \alpha_4\}
    \label{eq:agent_definition}
\end{equation}
where each agent $\alpha_i$ is responsible for generating artifact $A_i$. The transformation between artifacts is defined as:

\begin{equation}
    A_i = f_{\alpha_i}(A_{i-1}, U)
\label{eq:artifact_transformation}
\end{equation}
where $\mathcal{P}_i \subseteq \{A_1, \dots, A_{i-1}\}$ denotes the set of previously generated artifacts required at stage $i$, and $U$ represents user inputs (e.g., clarifications or constraints).

\begin{figure}
    \centering
    \fbox{\includegraphics[width=\linewidth]{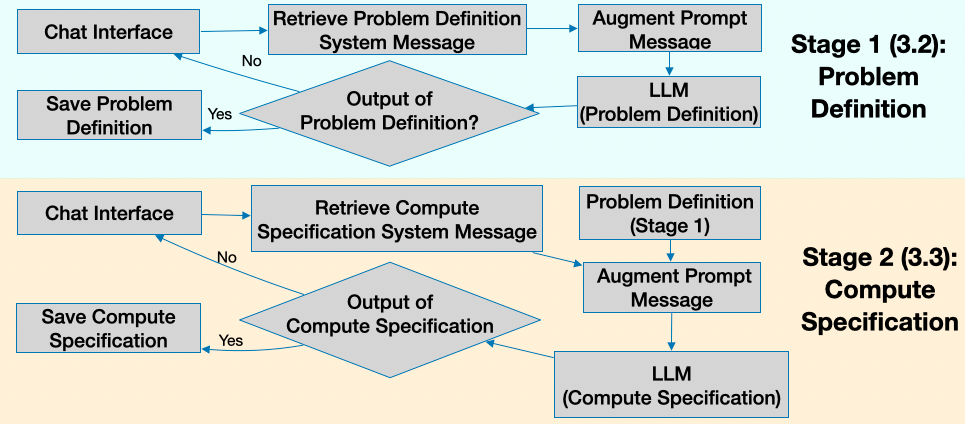}}
    \caption{DDAP Workflow (Stage 1 and 2)}
    \label{fig:ddap-workflow-1}
\end{figure}

\begin{figure}
    \centering
    \fbox{\includegraphics[width=0.9\linewidth]{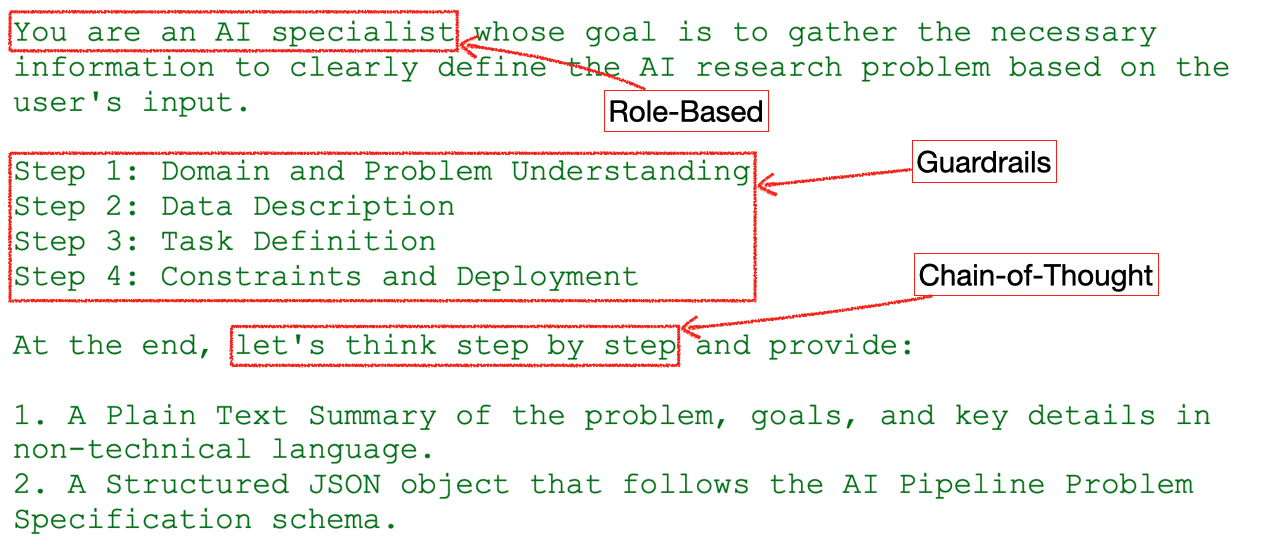}}
    \caption{Problem Definition System Message (Excerpt)}
    \label{fig:ddap-problem-definition}
\end{figure}

\subsection{Problem Definition (Stage 1)}\label{label:stage1}
The first stage of the framework focuses on eliciting and structuring the AI task into a semi-formal representation that can be consumed by subsequent stages. Rather than assuming a fully specified problem, \emph{DDAP} adopts a controlled approach that progressively guides the researcher in clarifying objectives, data characteristics, and constraints. This stage operationalizes the transition from high-level domain intent to a semi-structured problem definition. Unlike existing approaches that assume well-defined inputs~\cite{shyam2025bridging, maddireddy2025locoml}, \emph{DDAP} begins with an open-ended dialogue that incrementally refines user intent through guided interaction. The framework profiles the researcher’s domain and expertise level, and dynamically adapts its questioning strategy accordingly. For example, a medical researcher working on MRI-based tumour classification can be guided using biomedical terminology and clinically relevant constraints, whereas an agricultural researcher focusing on crop yield prediction can receive domain-specific prompts aligned with agronomic concepts. This adaptability is fostered by the LLM's capability of cross-domain reasoning, combined with role-based prompting and contextual augmentation.

The pipeline orchestrator coordinates this stage, as shown in Figure~\ref{fig:ddap-architecture}, by managing a controlled interaction loop. Specifically, the orchestrator manages the dialogue flow, augments prompts with contextual information, and evaluates the status of messages from the LLM, i.e., a response output or clarifying question, as detailed in Figure~\ref{fig:ddap-workflow-1}. At this stage, \emph{DDAP} integrates role-based prompting~\cite{shanahan2023role}, guardrail mechanisms~\cite{ayyamperumal2024current, rebedea2023nemo}, and chain-of-thought reasoning~\cite{wei2022chain} to foster structured and reliable interactions. Figure~\ref{fig:ddap-problem-definition} illustrates an excerpt of the system message that encodes these control and guidance mechanisms. Once a coherent problem definition is established, it is stored as a semi-structured artifact by the \emph{Output Handler}, which can be consumed by the downstream stages.

\begin{figure}
    \centering
    \fbox{\includegraphics[width=\linewidth]{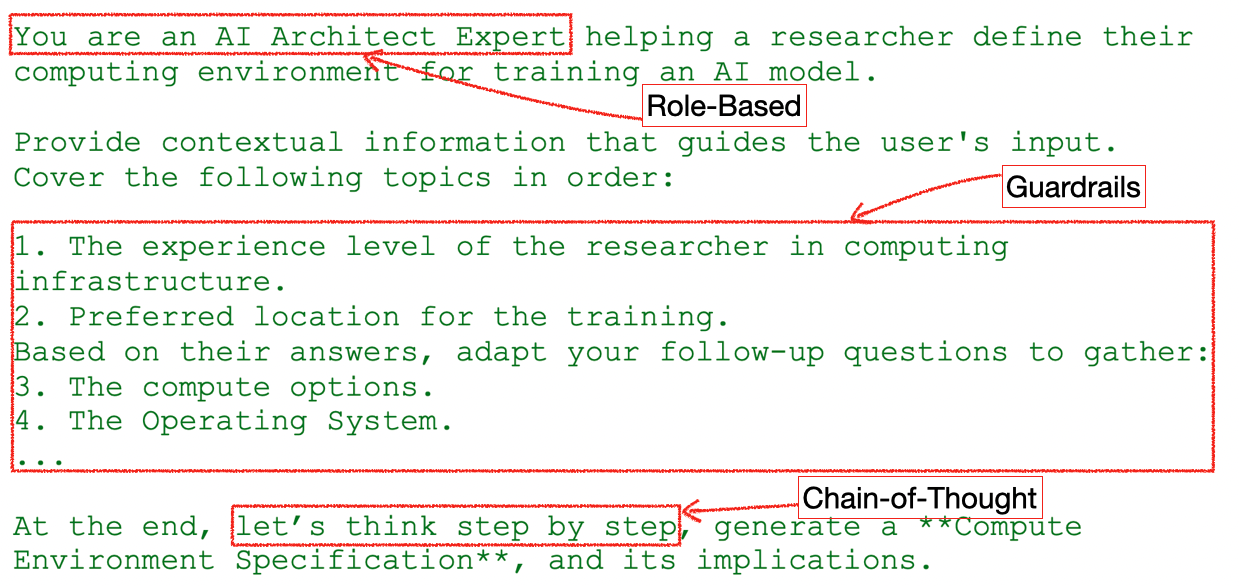}}
    \caption{Compute Environment Specification System Message (Excerpt)}
    \label{fig:ddap-compute-specification}
\end{figure}

\subsection{Compute Environment Specification (Stage 2)}\label{label:stage2}
The second stage focuses on capturing and formalizing the computing requirements necessary to implement the AI task. Building on the structured artifact produced during the \emph{Problem Definition} Stage, \emph{DDAP} elicits available compute environment specifications, including environment location (on-premises or cloud), accelerators (GPUs/TPUs), storage capacity, and budget constraints. This stage operationalizes the alignment between problem requirements and feasible computational resources. 
% \emph{DDAP} supports two complementary scenarios: \emph{known} and \emph{new} compute environments.

% In the \textbf{Known Compute Environment} scenario, users already have access to infrastructure (e.g., institutional clusters or cloud accounts). The framework guides users through targeted, structured interactions to capture relevant resource details, such as available hardware, storage limits, and access policies. In addition, users specify budgetary or usage constraints. The pipeline orchestrator then synthesizes this information to generate one or more candidate compute configurations, each accompanied by explicit trade-offs (e.g., cost vs. performance). This multi-option design aims to enable informed decision-making while maintaining alignment with the problem requirements.

% In the \textbf{New Compute Environment} scenario, users may lack prior experience with high-performance computing or cloud platforms. In this case, \emph{DDAP} adopts a more guided interaction strategy, eliciting high-level constraints such as budget, preferred deployment location, and expected workload characteristics. Based on this information, the framework proposes feasible compute configurations tailored to the task, abstracting low-level technical details while preserving essential decision criteria. This approach lowers the entry barrier for non-experts while still producing structured, machine-interpretable artifacts suitable for downstream stages.

The pipeline orchestrator (see Figure~\ref{fig:ddap-architecture}) controls the interaction loop, similar to the previous stage, as detailed in Figure~\ref{fig:ddap-workflow-1}. The orchestrator reuses the problem-definition artifact to contextualize prompts and maintain consistency across the first two stages. Role-based prompting and chain-of-thought reasoning are employed to guide the LLM in eliciting relevant information and reasoning about resource configurations, while guardrails maintain that essential factors, including compute location, accelerators, storage, and budget, are systematically addressed. Figure~\ref{fig:ddap-compute-specification} shows an excerpt of the system message used to encode these constraints. During the interaction, the orchestrator continuously evaluates the status, i.e., a question or output response, of the generated responses, guiding users toward the next expected action. Once a valid configuration is established, the resulting specification is stored as a structured artifact by the \emph{Output Handler}. The resulting compute environment specifications are reusable across projects; for example, a configuration defined for MRI-based tumour classification can be reused for other medical imaging tasks with similar computational requirements.

\subsection{Pipeline Generation (Stage 3)}\label{lbl:pipeline-stage}
The third stage of \emph{DDAP} focuses on generating structured AI pipeline specifications that can be translated into executable implementations, e.g., PyTorch or TensorFlow. This stage builds on the artifacts produced in the \emph{Problem Definition} and \emph{Compute Environment Specification} stages. To address the inherent complexity of AI pipeline design, \emph{DDAP} decomposes this stage into two sequential steps: (1) preprocessing techniques generation and (2) pipeline specification generation. This decomposition aims to improve more focused reasoning at each step and reduce the likelihood of errors in the final pipeline design.

\begin{figure}
    \centering
    \fbox{\includegraphics[width=\linewidth]{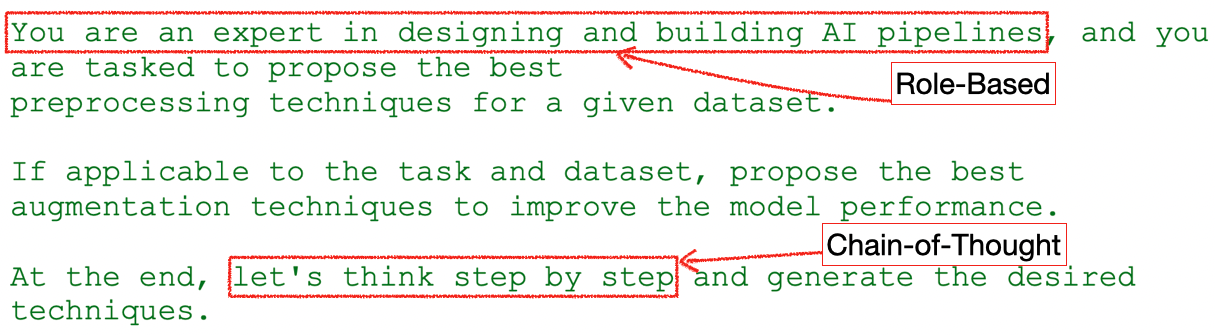}}
    \caption{Preprocessing Generation System Message (Excerpt)}
    \label{fig:ddap-preprocessing-message}
\end{figure}

\begin{figure}
    \centering
    \fbox{\includegraphics[width=\linewidth]{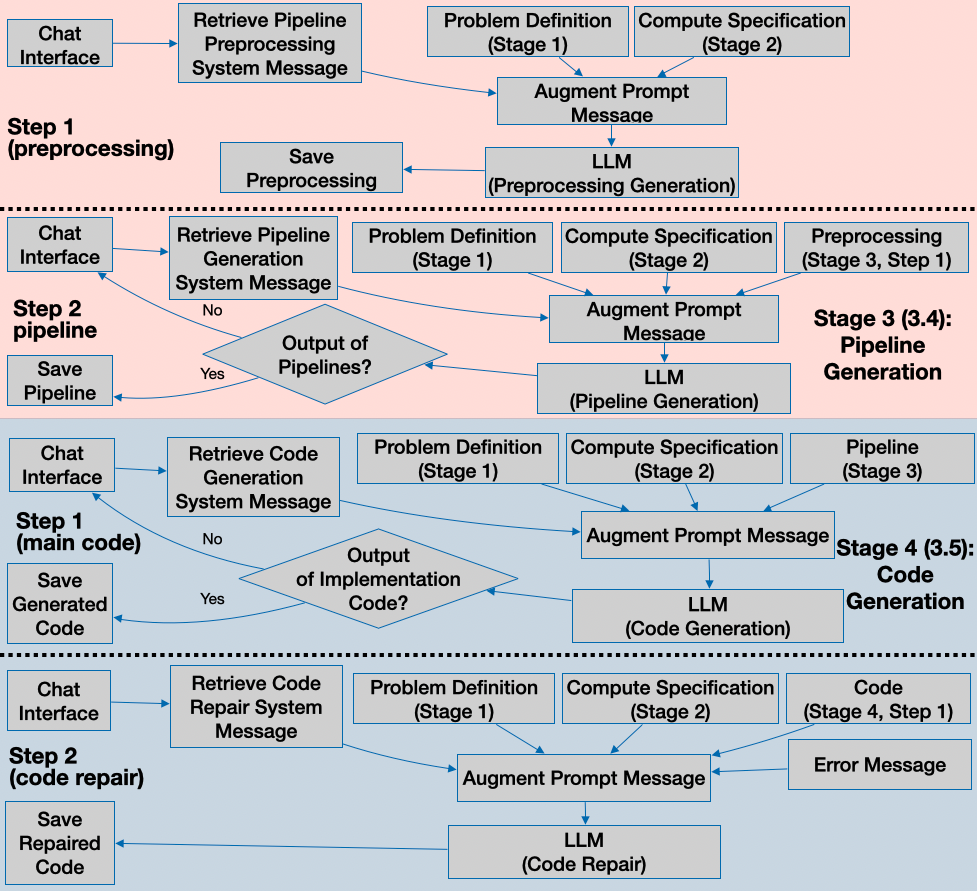}}
    \caption{DDAP Workflow (Stage 3 and 4)}
    \label{fig:ddap-workflow-2}
\end{figure}

\begin{figure}
    \centering
    \fbox{\includegraphics[width=\linewidth]{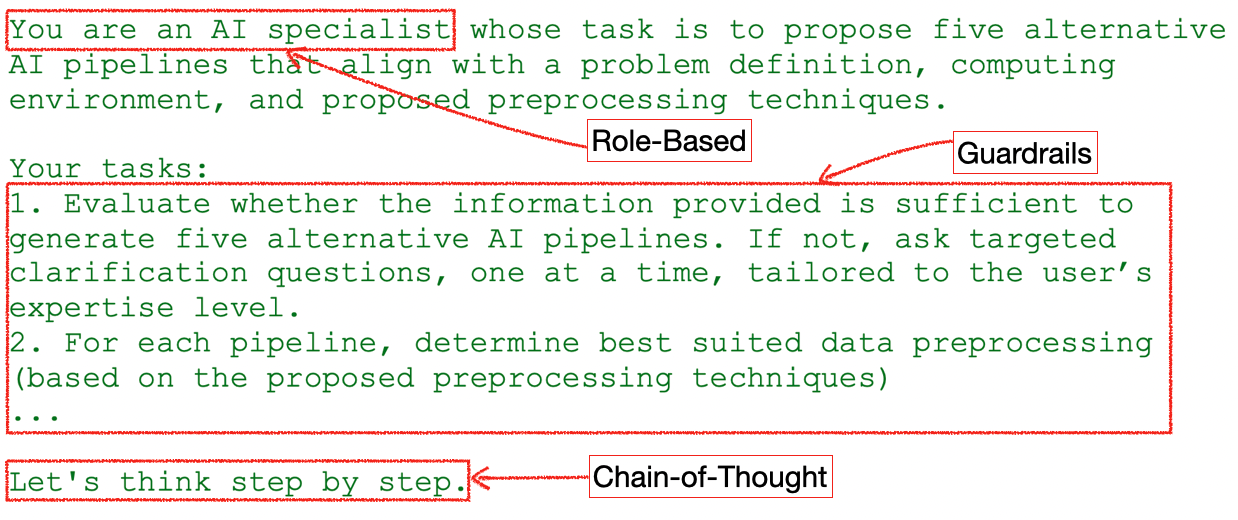}}
    \caption{Pipeline Generation System Message (Excerpt)}
    \label{fig:ddap-pipeline-specification-message}
\end{figure}

In the first step, a one-shot interaction is used to generate candidate preprocessing strategies tailored to the task and data characteristics, and the system message is shown in Figure~\ref{fig:ddap-preprocessing-message}. This step leverages the problem definition and compute environment specifications from the previous stages to capture domain-specific considerations such as feature engineering, normalization, handling missing values, and data transformations, as detailed in Figure~\ref{fig:ddap-workflow-2}. By explicitly separating preprocessing from the full pipeline generation, \emph{DDAP} aims to ensure that downstream decisions are grounded in a coherent data preparation strategy. In the second step, the pipeline orchestrator uses the generated preprocessing artifact, along with the problem definition and compute environment specification, to guide the generation of complete pipeline designs, as shown in Figure~\ref{fig:ddap-pipeline-specification-message}. This stage combines role-based prompting, guardrail mechanisms, chain-of-thought reasoning~\cite{wei2022chain}, and retrieval-augmented generation (RAG)~\cite{zhao2024retrieval, arslan2024survey} to streamline the interaction and improve the quality of outputs. Prior work has shown that decomposing complex tasks into multiple reasoning steps can significantly improve LLM performance and reduce errors~\cite{wei2022chain, yao2022react, yang2024multi}. \emph{DDAP} operationalizes this insight by structuring pipeline generation as a staged interaction rather than a single prompt.

Within this controlled setup, the LLM is assigned the role of an AI pipeline specialist. Guided by contextual artifacts and system-level constraints, the model generates multiple alternative pipeline specifications. In our implementation, \emph{DDAP} generates five candidate pipelines per task, and this design decision is motivated by the probabilistic nature of LLMs, which may produce suboptimal or erroneous outputs in a single generation. By producing multiple alternatives, the framework aims to increase the likelihood of identifying a high-performing and valid pipeline without requiring iterative manual tuning. This orchestrator-led human-in-the-loop approach aims to ensure that pipeline designs remain aligned with both domain requirements, resource constraints, and users' intents. Once sufficient information has been collected, structured pipeline alternative specifications are generated, each with its pros and cons, and persisted as reusable artifacts for subsequent code generation.   

\subsection{Code Generation (Stage 4)}
The final stage of the \emph{DDAP} framework focuses on translating selected pipeline specifications into executable implementations on target platforms such as PyTorch or TensorFlow. This stage bridges the gap between structured pipeline design and practical implementation, ensuring that generated solutions are both functional and aligned with user requirements. A detailed workflow of this stage is shown in Figure~\ref{fig:ddap-workflow-2}.

\begin{figure}
    \centering
    \fbox{\includegraphics[width=\linewidth]{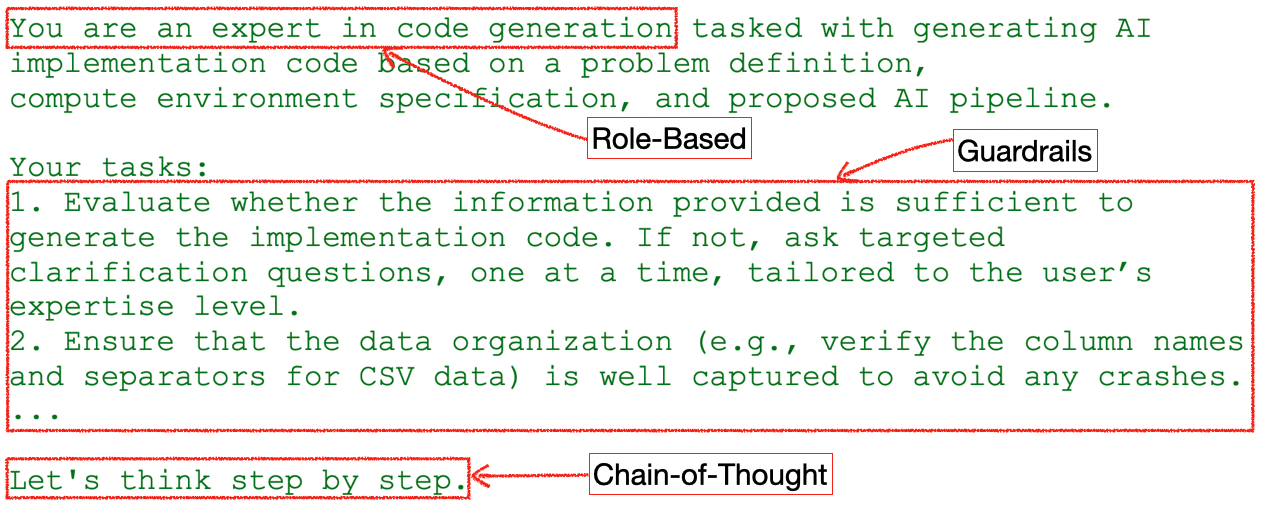}}
    \caption{Code Generation System Message (Excerpt)}
    \label{fig:ddap-code-generation-message}
\end{figure}

\begin{figure}
    \centering
    \fbox{\includegraphics[width=\linewidth]{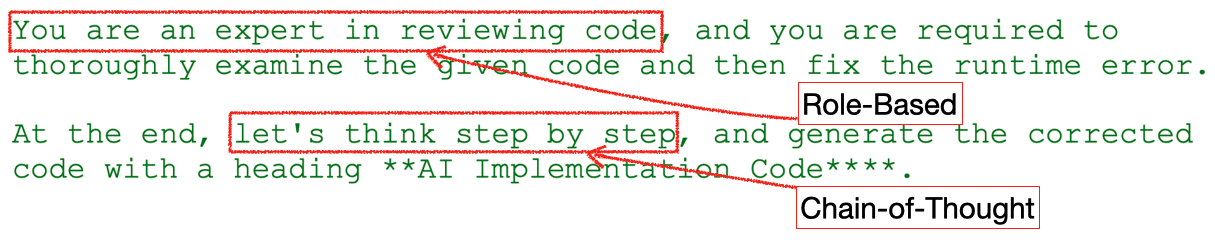}}
    \caption{Code Repair System Message (Excerpt)}
    \label{fig:ddap-code-repair-message}
\end{figure}

The pipeline orchestrator augments prompt messages with artifacts from previous stages, i.e., \emph{problem definition}, \emph{compute specification}, and \emph{pipeline specification}, as illustrated in Figure~\ref{fig:ddap-code-generation-message}. Formally, the overall orchestrator process can be viewed as a staged composition of transformations:
\begin{equation}
    A_4 = f_{\alpha_4} \circ f_{\alpha_3} \circ f_{\alpha_2} \circ f_{\alpha_1}(U_0)
    \label{eq:orchestrator}
\end{equation}
where $U_0$ is the initial user intent.

At this stage, Code-oriented LLMs, such as Code Llama~\cite{code-llama}, are then employed to generate executable code tailored to the defined compute environment and preferred platform. The orchestrator manages a controlled interaction loop, aiming to generate implementation code that is consistent with upstream design decisions and resource constraints. To improve correctness and alignment, and streamline the interaction steps, \emph{DDAP} combines role-based prompting, guardrail mechanisms, chain-of-thought reasoning~\cite{wei2022chain}, and retrieval-augmented generation (RAG). Here, the LLM is assigned the role of a code generation expert (Figure~\ref{fig:ddap-code-generation-message}), configuring it to produce complete, structured implementations.

Recognizing that LLM-generated code may contain syntactic or runtime errors due to the probabilistic nature of generation, \emph{DDAP} introduces an optional \emph{code repair} one-shot interaction. To this effect, the orchestrator incorporates code with an error and its error message into the prompt, allowing the model to identify and fix issues such as missing dependencies, incorrect configurations, or runtime failures, as illustrated in Figures~\ref{fig:ddap-workflow-2} and~\ref{fig:ddap-code-repair-message}. This repair mechanism improves the robustness and executability of the generated code with errors without requiring manual debugging from the user.

%% file: content/4-DDAP_Experiments.tex
\section{Evaluation}\label{sec:experiments}
We evaluate the applicability of \emph{DDAP} through a set of empirical studies derived from existing published work across multiple domains, including Biology, Business, Computer Science, and Health and Medicine. Rather than selecting benchmark datasets in isolation, we construct each experiment by pairing a published study with its corresponding dataset obtained from the University of California, Irvine (UCI) Machine Learning Repository~\cite{uci_dataset}. This setup allows a direct comparison between DDAP-generated pipelines and expert-designed solutions under comparable conditions.

For each study, we extract the problem formulation and other contextual information (e.g., dataset characterization and constraints) from the corresponding paper. These elements are then used for the natural language interaction with the \emph{DDAP} framework, simulating how a domain expert would specify the problem. The framework generates the associated artifacts, including the problem definition, compute environment specification, pipeline design, and executable code.
The generated pipelines are executed using the same datasets as the original studies. To ensure fair comparison, we adopt the evaluation metrics reported in each original study whenever possible, even when these metrics differ across tasks (e.g., classification, regression, and clustering). The performance of DDAP-generated models is then compared against the results reported in the corresponding studies.
%Overall, the results show that DDAP can generate pipelines that achieve competitive performance with respect to expert-designed models, despite operating in a single-pass setting without iterative hyperparameter tuning. However, performance degrades in more complex settings, such as textual clustering, highlighting current limitations and the need for improved representation learning and iterative refinement mechanisms.
Table~\ref{tab:experiments-results-comparison} presents the results of the six experiments conducted in this evaluation. The complete chat histories, structured artifacts (problem definition and compute specification), as well as the generated pipelines and implementation code, are available in our public repository (replication package)~\cite{anonymous_2026_19241799}.

% The complete chat histories, structured artifacts (problem definition and compute specification), as well as the generated pipelines and implementation code, are available in our public repository (replication package)~\footnote{https://zenodo.org/records/19241799}

% ~\cite{anonymous_2026_19241799}.

\begin{table}[t]
\centering
\caption{Comparison between \emph{DDAP} framework and reported paper results.}
\label{tab:experiments-results-comparison}
\small
\renewcommand{\arraystretch}{1.2}
\resizebox{\columnwidth}{!}{
\begin{tabular}{|l|l|l |l| c| c|}
\hline
\textbf{Dataset} & \textbf{Data Type} & \textbf{Task} & \textbf{Metric} & \textbf{Our Result} & \textbf{Paper Result} \\
\hline

Jute Pest (\ref{jute-pests})  & Image & Classification & Precision & 0.95 & 0.97 \\
                     & & & Recall & 0.94 & 0.97 \\
                     & & & F1 Score & 0.94 & 0.97 \\
\hline

Parkinsons & Integer, & Regression & MAE (Motor                                                         UPDRS) & 0.30 & 4.50 \\
                    Telemonitoring (\ref{lbl:parkins}) & Real & & MAE (Total UPDRS) & 2.66 & 6.00 \\
\hline

Product (\ref{lbl:product}) & Text & Classification & Training Accuracy & 99.56\% & -- \\
                                            & & & Validation Accuracy & 96.83\% & 95\% \\
                                            & & & Precision & 0.97 & -- \\
                                            & & & Recall & 0.97 & -- \\
                                            & & & F1 Score & 0.97 & -- \\
\hline

Product (\ref{lbl:product}) & Text & Clustering & Accuracy & 0.00085 & -- \\
              & & & Precision & 0.0015 & -- \\
              & & & Recall & 0.00085 & -- \\
              & & & F1 Score & 0.0010 & 0.59 \\
\hline

Customer & Integer, & Clustering & Silhouette Score & 0.53 & Manual \\
                        Segmentation (\ref{lbl:customer_segmentation}) & Real & & & & Analysis \\
                          
\hline

Real Waste (\ref{lbl:real_waste}) & Image & Classification & Precision & 0.85 & 0.91 \\
                      & & & Recall & 0.84 & 0.88 \\
                      & & & F1 Score & 0.84 & 0.90 \\
\hline

Stock Market (\ref{lbl:stock_forecast}) & Real & Regression & Validation Accuracy & 0.79 & 0.68 \\
                    & & & Precision & 0.78 & -- \\
                    & & & Recall & 0.88 & -- \\
                    & & & F1 Score & 0.83 & -- \\
\hline
\end{tabular}}
\end{table}

\subsection{Jute Pests Classification} \label{jute-pests}
The first experiment illustrates how a researcher in the Biology domain can use \emph{DDAP} to develop an AI solution for classifying jute pests into 17 categories. Jute is an important crop globally, and pest infestation is a major factor affecting its yield. This classification model aims to enhance timely intervention to reduce financial losses and improve crop yield. We use the Jute Pest dataset~\cite{jute_pest_920}, which comprises 7,235 images. Through natural language interaction and by following the four stages described in Section~\ref{framework-architecture}, the research goal is defined as a multi-class classification task over jute pest categories. The \emph{DDAP} framework dynamically adapts follow-up questions based on the domain context and the user’s level of expertise.

\begin{figure}
    \centering
    \fbox{\includegraphics[width=\linewidth]{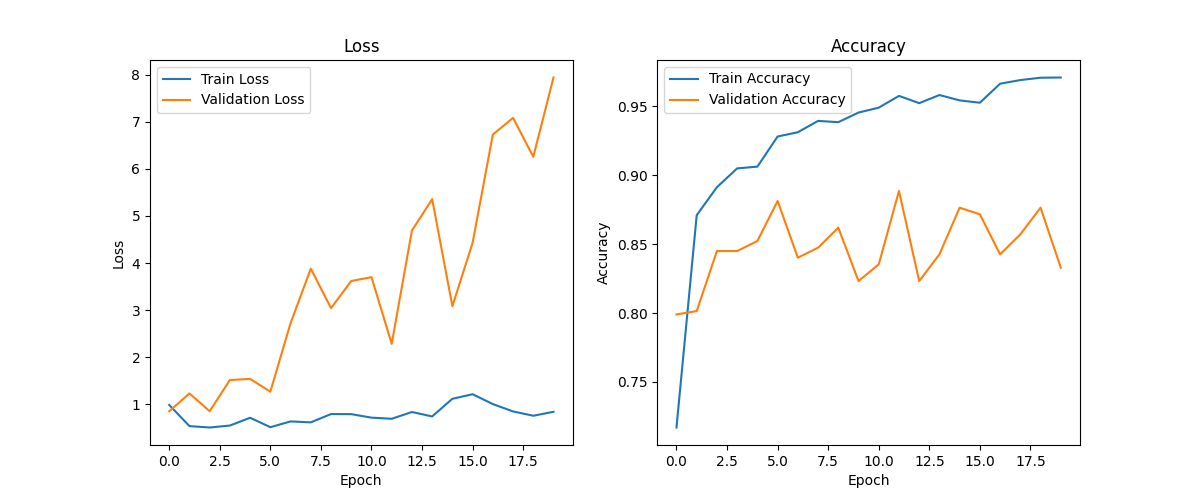}}
    \caption{Jute Pests Classification Model Performance}
    \label{fig:jute-pests-classification}
\end{figure}

In this experiment, \emph{DDAP} generates the following pipeline: (1) Basic Convolutional Neural Network (CNN), (2)
Transfer Learning with Pre-trained Model, (3) Transfer Learning with Fine-tuning, (4) Ensemble Learning with Multiple Models, and (5) Custom CNN with Advanced Augmentation Techniques. The Transfer Learning algorithm (Pipeline 2) outperforms others, achieving a validation accuracy of 83.29\%, precision of 0.95, recall of 0.94, and F1-score of 0.94, and the loss and accuracy graphs are shown in Figure~\ref{fig:jute-pests-classification}. In comparison, the best-performing model reported in the original study~\cite{islam2024efficient} achieves a precision, recall, and F1-score of 0.97 each. Although the \emph{DDAP}-generated model performs slightly below the expert-developed model, it demonstrates the ability to design and train effective AI solutions entirely through natural language interaction in a single pass of the \emph{DDAP} workflow, without iterative optimization or manual hyperparameter tuning.

\begin{figure}
    \centering
    \fbox{\includegraphics[width=0.6\linewidth]{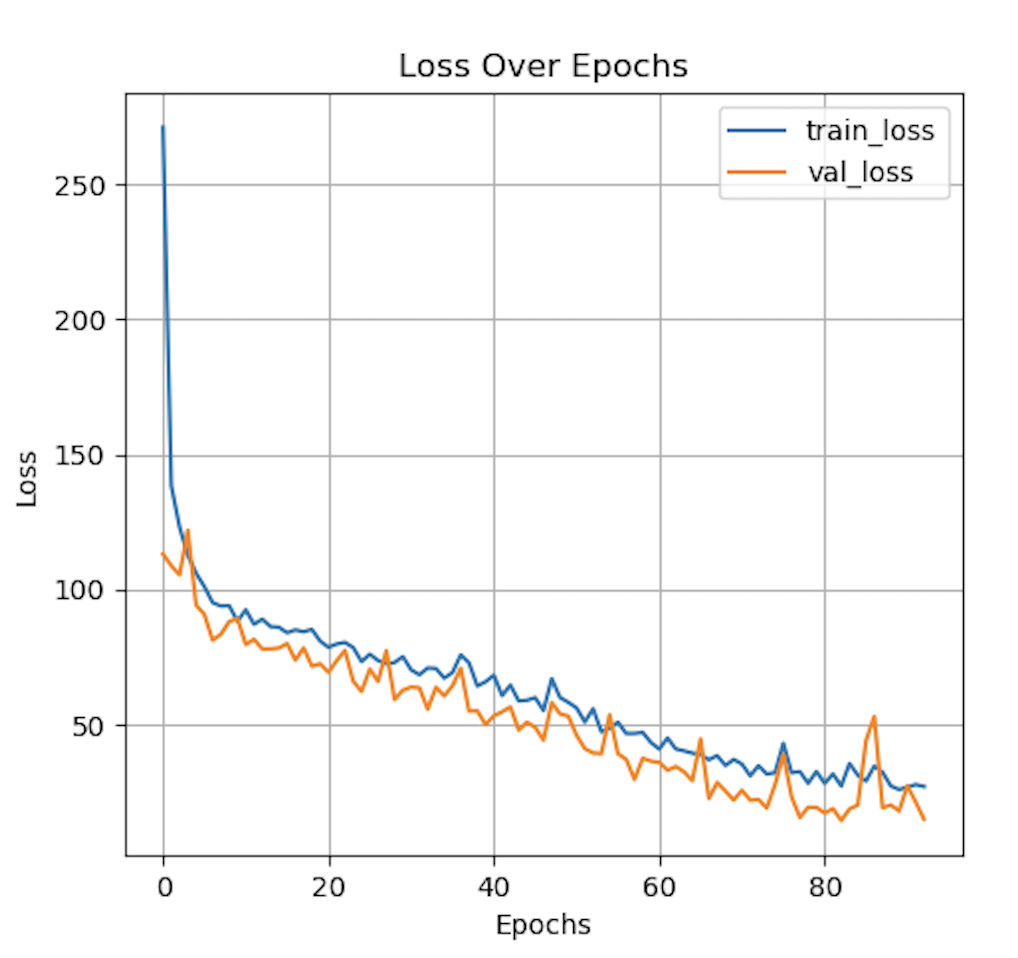}}
    \caption{Parkinsons Telemonitoring Model Performance}
    \label{fig:parkins-loss-plot}
\end{figure}

\subsection{Parkinsons Telemonitoring} \label{lbl:parkins}
The second experiment illustrates how a researcher in the medical domain can use \emph{DDAP} to design and develop an AI model from scratch through natural language interaction. The objective  of this experiment is to remotely monitor the progression of Parkinson’s Disease (PD) by analyzing patients’ speech signals, as an alternative to clinical visits and physical examinations, which are costly and time-consuming~\cite{tsanas2009accurate}. Specifically, the task is to estimate the Unified Parkinson’s Disease Rating Scale (UPDRS)~\cite{tsanas2009accurate} from speech recordings. We use the Parkinson's Telemonitoring dataset~\cite{parkinsons_telemonitoring_189}, which comprises 5,875 records, to train a model that predicts PD severity based on voice features.

Following the four-stage \emph{DDAP} workflow, our framework generates five alternative pipelines: (1) Random Forest, (2) Gradient Boosting Machines (GBM), (3) Long Short-Term Memory (LSTM), (4) Convolutional Neural Network (CNN), and (5) an Ensemble model. We evaluate the models using Mean Absolute Error (MAE) on both Motor UPDRS and Total UPDRS. Among the generated pipelines, the Ensemble model (Pipeline 5) achieves the best performance, with MAE values of 0.30 (Motor UPDRS) and 2.66 (Total UPDRS). The loss curves are shown in Figure~\ref{fig:parkins-loss-plot}. In comparison with the original study~\cite{tsanas2009accurate}, the expert-developed model achieves MAE values of 4.5 and 6.0 for Motor and Total UPDRS, respectively. These results indicate that, even without iterative optimization and manual hyperparameter tuning, \emph{DDAP} can generate models that outperform expert-designed approaches.

\subsection{Product Classification and Clustering}\label{lbl:product}
The rapid expansion of the e-commerce sector has made product retrieval increasingly critical. As businesses transition online, the volume and diversity of product information grow substantially, making it more challenging for users to effectively select and compare identical products. To address this challenge, this experiment illustrates how a researcher in the business domain can leverage the \emph{DDAP} framework to design and build an AI model for classifying e-commerce products based on their titles. We use the Product Classification and Clustering dataset~\cite{product_classification_and_clustering_837}, which comprises 35,311 tabular records, including product titles, vendors, and categories, to train a classification model.

\begin{figure}
    \centering
    \fbox{\includegraphics[width=0.95\linewidth]{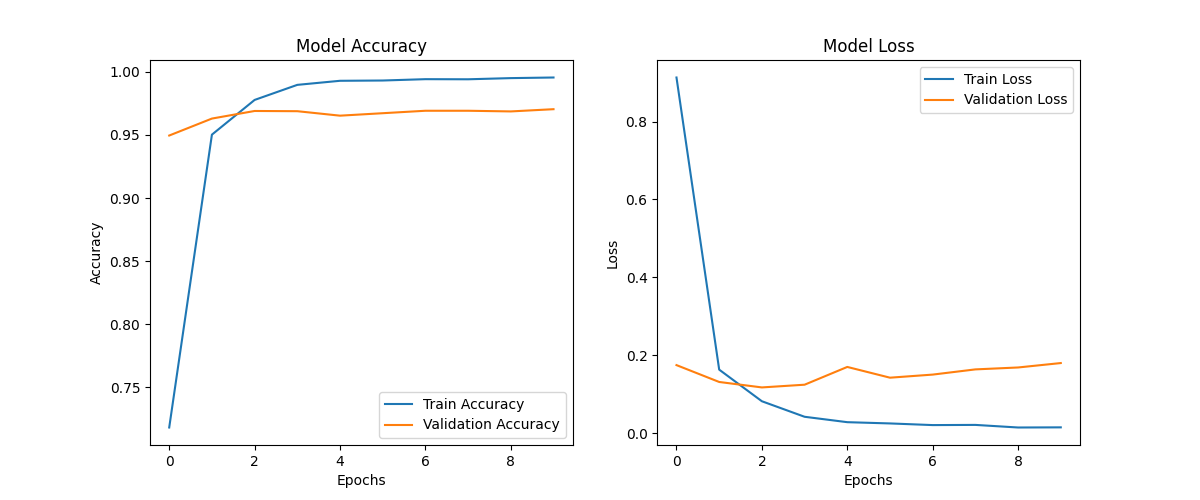}}
    \caption{Product Classification Model Performance}
    \label{fig:product-classification-plot}
\end{figure}

Following the four-stage \emph{DDAP} workflow, our framework generates five alternative pipelines: (1) Feedforward Neural Network, (2) Convolutional Neural Network (CNN) for text classification, (3) Recurrent Neural Network (RNN) with LSTM, (4) Transformer-based model, and (5) Ensemble learning with neural networks. Among these, the CNN-based pipeline (Pipeline 2) achieves the best performance, with a training accuracy of 99.56\%, validation accuracy of 96.83\%, precision of 0.97, recall of 0.97, and F1-score of 0.97. In comparison with the original work~\cite{akritidis2018effective}, the expert-developed model achieves an accuracy of 95\%. Hence, the \emph{DDAP} model achieves superior performance despite being generated solely through natural language interaction and without iterative optimization. The training accuracy and loss curves are shown in Figure~\ref{fig:product-classification-plot}.

In addition, we evaluate \emph{DDAP} on a clustering task using the same dataset, where products are grouped based on their titles. Our framework generates five pipelines: (1) Density-Based Spatial Clustering of Applications with Noise (DBSCAN) with Term Frequency–Inverse Document Frequency (TF-IDF) Vectorization, (2) Hierarchical Clustering with word embeddings, (3) Gaussian Mixture Models (GMM) with TF-IDF, (4) Spectral Clustering with Principal Component Analysis (PCA), and (5) Affinity Propagation with data augmentation. However, all pipelines perform poorly in this setting. The best-performing pipeline, Affinity Propagation, achieves very low scores (accuracy: 0.00085, precision: 0.0015, recall: 0.00085, and F1-score: 0.001), which are significantly below the expert-reported performance (F1-score of 0.59)~\cite{akritidis2020self}. These results highlight a key limitation of the current \emph{DDAP} framework in handling textual clustering tasks, suggesting the need for more advanced representation learning, task-specific preprocessing, and iterative optimization strategies in future work.

\begin{figure}
    \centering
    \fbox{\includegraphics[width=0.8\linewidth]{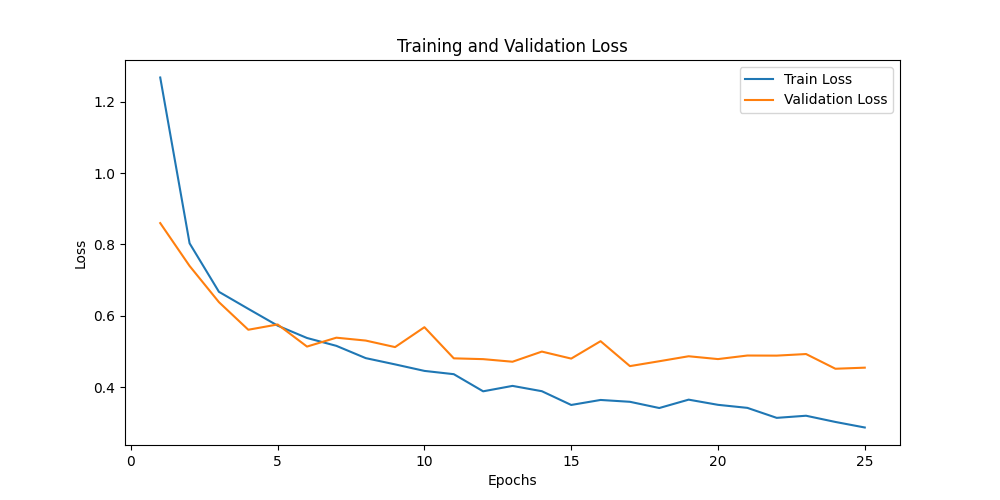}}
    \caption{Waste Material Classification Model Performance}
    \label{fig:waste-material-classification-plot}
\end{figure}

\begin{table*}
\centering
\caption{Comparison of Customer Segmentation Clustering Results Between Expert-Driven Methods and DDAP}
\label{tab:online_customers_metrics_comparison}
\renewcommand{\arraystretch}{1.2}
\begin{tabular}{|c|ccc|c||c|ccc|c|}
\hline
\multicolumn{5}{|c||}{\textbf{Expert-Driven~\cite{chen2012data}}} 
& \multicolumn{5}{c|}{\textbf{\emph{DDAP} Framework}} \\
\hline
\textbf{Cluster} & \textbf{Percentage} & \textbf{Recency} & \textbf{Frequency} & \textbf{Monetary} 
& \textbf{Cluster} & \textbf{Percentage} & \textbf{Recency} & \textbf{Frequency} & \textbf{Monetary} \\
\hline

1 & 14.14 & 9.8 & 1.3 & 361.20 
& 1 & 70.10 & 42.0 & 41.0 & 121.85 \\

2 & 17.07 & 5.4 & 2.3 & 586.19 
& 2 & 17.84 & 270.0 & 17.0 & 58.91 \\

3 & 46.91 & 1.5 & 2.6 & 685.71 
& 3 & 11.71 & 12.0 & 274.0 & 800.99 \\

4 & 16.83 & 1.0 & 8.3 & 2425.09 
& 4 & 0.28 & 5.5 & 1557.0 & 5277.37 \\

5 & 5.05 & 0.7 & 17.7 & 5962.85 
& 5 & 0.09 & 1.5 & 5391.5 & 22811.02 \\

\hline
\multicolumn{5}{|c||}{\textbf{Silhouette Score: Not Reported}} 
& \multicolumn{5}{c|}{\textbf{Silhouette Score: 0.5297}} \\
\hline
\end{tabular}
\end{table*}

\subsection{Customer Segmentation}\label{lbl:customer_segmentation}
The rapid growth of online retail, coupled with the ability to capture detailed customer interaction data, has transformed how businesses understand and engage with customers. To address key business objectives, such as identifying high-value customers, understanding purchase behaviour, and predicting responses to marketing campaigns, organizations increasingly rely on data-driven techniques. In this experiment, we design and build a clustering-based machine learning model to segment customers into meaningful groups, aiming for more effective customer-centric marketing strategies. We use the Online Retail dataset~\cite{online_retail_352}, which contains 541,909 transaction records from a UK-based online retail store between 01/12/2010 and 09/12/2011. To train the clustering model, we employ the \emph{DDAP} framework to generate the required artifacts, including five alternative pipelines: (1) K-Means Clustering with standardization, (2) DBSCAN, (3) Hierarchical Clustering, (4) Gaussian Mixture Model (GMM), and (5) Spectral Clustering. Among the pipelines, the Hierarchical Clustering pipeline (Pipeline 3) achieves the best performance, with a silhouette score of 0.53, which indicates well-separated clusters and reveals a Pareto-like distribution~\cite{koch201180}. Also, we adopt metrics consistent with the expert-driven study~\cite{chen2012data}, and report a comparison between the \emph{DDAP}-generated results and the expert-designed models in Table~\ref{tab:online_customers_metrics_comparison}.

These results demonstrate that the proposed human-in-the-loop agentic framework can generate meaningful RFM-based customer segments~\cite{chen2012data} using only natural language input, without iterative manual tuning or expert intervention. The resulting clusters exhibit clear behavioral distinctions aligned with classical RFM analysis, including average (Cluster 1), low-value (Cluster 2), high-value (Cluster 3), and very high-value (Clusters 4 and 5) customer groups, even with the absence of explicit preprocessing steps such as outlier removal. Although the segmentation is less balanced than expert-designed models, the overall customer hierarchy and derived insights remain consistent, highlighting the potential of \emph{DDAP} in democratizing data-driven customer analytics.

\subsection{Waste Material Classification}\label{lbl:real_waste}
This experiment examines how non-experts can design and train machine learning models to classify waste materials into different categories, such as recyclable inorganics, divertible organics, and non-recyclable inorganics~\cite{single2023realwaste}. Improper waste management and disposal pose significant risks to environmental sustainability and human health~\cite{oman2008chemical}. To address this challenge, we leverage the \emph{DDAP} framework using the \emph{RealWaste} dataset~\cite{realwaste_908}, which comprises 4,752 images, to train classification models.

Following the \emph{DDAP} workflow, the framework generates five alternative pipelines: (1) Basic CNN, (2) Transfer Learning with a pre-trained model (e.g., ResNet50), (3) CNN with data augmentation, (4) Ensemble learning with multiple models, and (5) Custom CNN with advanced augmentation techniques. We evaluate the performance across these pipelines and observe that the Transfer Learning approach (Pipeline 2) achieves the best results, with an accuracy of 85\%, precision of 0.84, recall of 0.84, and F1-score of 0.84 on the test set. The training and validation curves are shown in Figure~\ref{fig:waste-material-classification-plot}. In comparison with the original study~\cite{single2023realwaste}, expert-designed models achieve up to 89.19\% accuracy using carefully tuned architectures such as InceptionV3. Although the \emph{DDAP}-generated model performs slightly lower, the results remain competitive despite the absence of hyperparameter tuning or iterative optimization. 

\subsection{Stock Market Forecast}\label{lbl:stock_forecast}
This experiment further illustrates how a non-expert can leverage the \emph{DDAP} framework to build a predictive model for forecasting stock market direction. We use the real-world \emph{ISTANBUL STOCK EXCHANGE} time-series dataset~\cite{istanbul_stock_exchange_247}, which comprises 536 records and includes returns from the Istanbul Stock Exchange along with several international indices.

\begin{figure}
    \centering
    \fbox{\includegraphics[width=\linewidth]{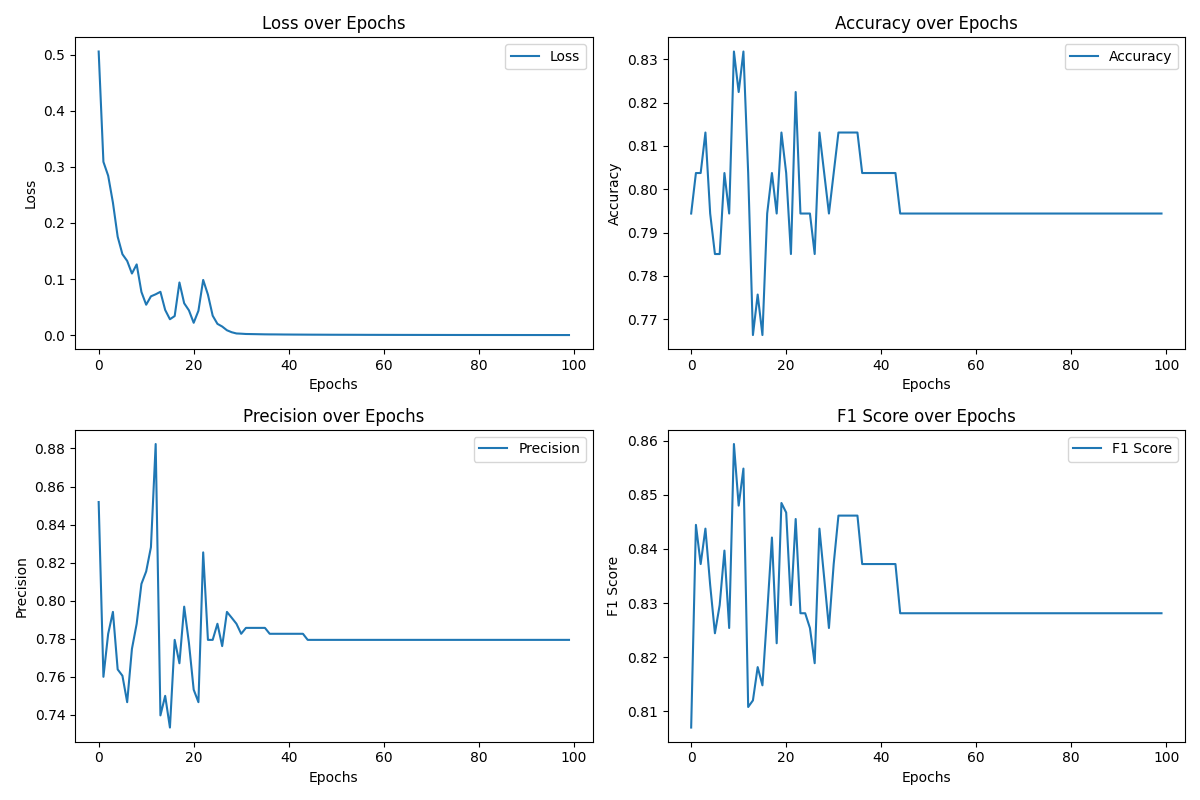}}
    \caption{Stock Market Forecast Model Performance}
    \label{fig:stock-forecast-plot}
\end{figure}

Following the four-stage \emph{DDAP} workflow, \emph{DDAP} framework generates five alternative pipelines: (1) Long Short-Term Memory (LSTM), (2) Gated Recurrent Unit (GRU), (3) Transformer-based model, (4) Convolutional Neural Network (CNN) for time series, and (5) an ensemble method combining LSTM and CNN. Among these, the Transformer-based pipeline achieves the best performance, with an accuracy of 79.44\%, precision of 0.78, recall of 0.88, and F1-score of 0.83. The training curves (loss, accuracy, precision, and f1-score) are depicted in Figure~\ref{fig:stock-forecast-plot}.

In comparison with the original work~\cite{akbilgic2014novel}, the expert-designed model achieves up to 68\% accuracy. Hence, the \emph{DDAP}-generated model demonstrates good predictive performance even without the iterative optimization and hyperparameter tuning process. Moreover, unlike the original study, which primarily reports directional accuracy, our framework provides a more comprehensive evaluation using precision, recall, and F1-score, offering deeper insight into model behavior.

In conclusion, the experimental results summarized in Table~\ref{tab:experiments-results-comparison} demonstrate that the \emph{DDAP} framework can generate competitive AI models across different domains, including image classification, regression, text classification, and clustering. In most classification and regression tasks, the performance of \emph{DDAP}-generated models is competitive, although it varies significantly in more complex tasks such as textual clustering. For example, in the Parkinson’s telemonitoring and stock market forecasting tasks, the framework achieves superior performance despite relying on a single-pass pipeline generation process without iterative optimization or hyperparameter tuning. Similarly, for product classification, \emph{DDAP} outperforms reported original work accuracy while maintaining high precision and recall. However, Table~\ref{tab:experiments-results-comparison} also highlights limitations in more complex tasks such as textual clustering, where the generated pipelines significantly underperform compared to expert approaches, indicating the need for improved representation learning and iterative refinement mechanisms.

%% file: content/5-Related_Work.tex
\section{Related Work}\label{sec:related}
This section positions the \emph{DDAP} framework against related work across three areas: (1) low-code and no-code (LCNC) framework for AI development, (2) automated machine learning (AutoML) and pipeline generation, and (3) LLM-based approaches for adaptive AI systems. We briefly review each stream and highlight how \emph{DDAP} advances the state of the art. 

\subsection{Low-Code and No-Code AI Frameworks}
Low-code and no-code (LCNC) frameworks democratize application and AI development by reducing programming requirements through visual programming and drag-and-drop interfaces~\cite{paliwal2024low, bock2021low}. Recent surveys reveal that LLM-based LCNC tools accelerate research by automating code generation, integrating AutoML components, and streamlining repetitive tasks. However, critical steps such as problem definition and training dataset design still require significant human effort~\cite{martinez2023using, albu2025ai}, underscoring the need for stronger guidance and domain-aware support. Platforms such as \emph{No-Code ML} and other web-based solutions~\cite{shyam2025bridging, maddireddy2025locoml} allow non-programmers to build ML models, while generative AI further expands capabilities through template generation, natural language coding, and adaptive workflows in domains such as healthcare and finance~\cite{paliwal2024low, kumar2023natural}. Despite these advances, current LCNC systems still face challenges in adapting to diverse user expertise levels, offering fine-grained control, and supporting artifact reuse~\cite{albu2025ai, d2024novel, shyam2025bridging}.

\subsection{Automated ML and Pipeline Generation}
Automated ML (AutoML) frameworks aim to lower expertise requirements by automating tasks such as preprocessing, model selection, and hyperparameter tuning~\cite{bergstra2011algorithms, snoek2012practical}. Systems such as Alpine Meadow~\cite{shang2019democratizing} integrate query optimization with bandit strategies for interactive pipeline generation, while DeepLine employs deep reinforcement learning to efficiently explore large search spaces~\cite{heffetz2020deepline}. The Data Feature and Service Association (DFSR) approach~\cite{ru2020machine} combines dataset characteristics with service association mining and Monte Carlo Tree Search~\cite{browne2012mcts} to rapidly construct candidate pipelines. SapientML~\cite{saha2022sapientml} leverages thousands of Kaggle notebooks to learn dataset–pipeline associations and then constructs pipelines through program synthesis. However, pipeline optimization has been shown to exhibit multiple local optima and high irregularity, making naive search strategies prone to failure~\cite{garciarena2018analysis}. This underscores the need for approaches that go beyond brute-force search by guiding pipeline design in a structured and adaptive manner. 

\subsection{LLM-Driven AI Development}
LLMs have extended automation beyond isolated tasks to cover end-to-end workflows. Tools such as GitHub Copilot~\cite{GitHubCopilot} and Cursor~\cite{Cursor} provide code-level assistance but still require extensive manual oversight~\cite{sonkin2025beyond}. Workflow-centric approaches like Prompt Pipeline Language introduce mechanisms for structuring multi-step tasks with testing and rollback~\cite{sonkin2025beyond}, while LAMBDA positions LLMs as orchestrators of heterogeneous data processing workflows~\cite{sun2025lambda}. Similarly, the Acumos framework illustrates pipeline-centric orchestration for composite models~\cite{panchal2023mlops}. Despite these advances, empirical studies reveal persistent issues with reliability and test quality, for instance, nearly half of Copilot’s generated unit tests fail~\cite{sonkin2025beyond}. To mitigate such shortcomings, recent research has explored self-assessment, automated test generation, chain-of-thought prompting, and guardrails~\cite{wei2022chain, ayyamperumal2024current}. These directions move toward more structured and reusable workflows but still lack systematic mechanisms for artifact management, reproducibility, and adaptation to user expertise, as well as domain-specific constraints.   
\subsection{Discussion}
These three streams reveal complementary strengths and shared gaps. LCNC frameworks improve accessibility but lack depth for complex AI tasks; AutoML automates pipelines but assumes predefined problems; and LLM-based tools provide adaptability but often yield ad hoc, fragile outputs. Our \emph{DDAP} framework is positioned to unify these directions by combining: (1) natural language accessibility from LCNC, (2) structured pipeline generation and optimization from AutoML, and (3) contextual adaptability from LLMs. Unlike existing approaches, \emph{DDAP} introduces reusable artifacts at every stage, problem definition, compute environment specification, natural language pipeline design, and code generation, which aims to produce outputs that are not only tailored to individual users but also reproducible and transferable across domains. Our aim is to advance the state of the art by bridging accessibility, flexibility, automation, and adaptability in a single domain-driven framework.

%% file: content/6-Conclusion.tex
\section{Conclusion}\label{sec:conclusion}
AI has become a cornerstone of contemporary research, with applications spanning healthcare, natural sciences, and social sciences. Researchers increasingly rely on AI pipelines to analyze complex datasets, automate experimentation, and accelerate discovery. This growing adoption highlights the need for frameworks that make AI development more accessible, adaptable, and reproducible across disciplines. In this paper, we propose Domain-Driven Adaptable AI Pipelines (\emph{DDAP}), a controlled, human-in-the-loop, agentic framework that guides researchers in transforming domain-specific goals into executable AI solutions through four interrelated stages: \emph{problem definition}, \emph{compute specification}, \emph{pipeline generation}, and \emph{code generation}. Through multiple empirical studies, we demonstrate that \emph{DDAP} can adapt to diverse problem contexts while producing structured, reusable artifacts and achieving competitive performance relative to expert-developed models. 

Compared to existing low-code/no-code frameworks, AutoML systems, and LLM-based approaches, \emph{DDAP} is designed to integrate accessibility, adaptability, and reproducibility within a single framework. By lowering entry barriers and embedding domain knowledge directly into pipeline design, our framework aims to contribute towards the democratization of AI development. Future work will focus on extending \emph{DDAP} with enhanced agentic capabilities, including mechanisms to support the partial or full automation of \emph{pipeline generation} and \emph{code generation} with minimal human intervention. Also, we plan to introduce a dedicated optimization agent that can iteratively improve model behaviour by automatically adjusting relevant hyperparameter values. \\

% \section*{Data Availability Statement}
% An anonymous replication package is available at~\cite{anonymous_2026_19241799}.